\def\year{2021}\relax
\documentclass[letterpaper]{article} 
\usepackage{paralist}
\usepackage{aaai212}  
\usepackage{times}  
\usepackage{helvet} 
\usepackage{courier}  
\usepackage[hyphens]{url}  
\usepackage{graphicx} 
\usepackage{natbib}  
\usepackage{caption} 
\usepackage{xspace}
\usepackage{booktabs}
\usepackage{amsmath}

\newcommand{\xhdr}[1]{\vspace{2mm} \noindent\textbf{#1}}
\newcommand{\datasetname}{YouNiverse\xspace}
\newcommand{\numchannelsapprox}{136k\xspace}
\newcommand{\numvideosapprox}{72.9M\xspace}
\newcommand{\YT}{YouTube\xspace}
\newcommand{\numchannelsexact}{136,470\xspace}
\newcommand{\numvideosexact}{72,924,794\xspace}
\newcommand{\numchannelsraw}{156,978\xspace}
\newcommand{\numvideosraw}{84M\xspace}
\newcommand{\SB}{socialblade.com\xspace}
\newcommand{\CC}{channelcrawler.com\xspace}
\newcommand{\numcommentsvideos}{20.5M\xspace} 
\newcommand{\numcomments}{8.6B\xspace}
\newcommand{\numcommentsraw}{10.3B\xspace}
\newcommand{\numuserscomments}{449M\xspace} 

\usepackage{hyphenat}
\usepackage{mathptmx}

\title{YouNiverse: Large-Scale Channel and Video Metadata \\from English-Speaking YouTube\thanks{This paper has been accepted at the 15th International Conference on Web and Social Media (ICWSM), please cite accordingly.}}

\usepackage{minibox}
\newcommand{\affilSize}{10pt}
\newcommand{\authorbox}[3]{
\minibox[c]{
\vspace{-1mm}
#1\\
\vspace{-1mm}
{\normalfont \fontsize{\affilSize}{\affilSize}\selectfont{}#2}\\
{\normalfont \fontsize{\affilSize}{\affilSize}\selectfont{}#3}
}
}

\author{
\authorbox{Manoel Horta Ribeiro}{EPFL}{manoel.hortaribeiro@epfl.ch}
\hspace{35mm}
\authorbox{Robert West}{EPFL}{robert.west@epfl.ch}
}

\begin{document}

\maketitle

\begin{abstract}
\YT plays a key role in entertaining and informing people around the globe. 
However, studying the platform is difficult due to the lack of randomly sampled data and of systematic ways to query the platform's colossal catalog.
In this paper, we present \datasetname, a large collection of channel and video metadata from English\hyp language \YT.
\datasetname comprises metadata for over \numchannelsapprox channels and \numvideosapprox videos published between May 2005 and October 2019, as well as channel-level time-series data of weekly subscriber and view counts.
Leveraging channel ranks from \SB, an online service that provides information about \YT, we are able to assess and enhance the representativeness of the sample of channels.
Additionally, the dataset also contains
a table specifying which videos a set of \numuserscomments anonymous users commented on.
\datasetname, publicly available at \url{https://doi.org/10.5281/zenodo.4650046},
will empower the community to do research with and about \YT.

\end{abstract}

\section{Introduction}

\YT plays an important role in society.
In 2018, 54\% of adult U.S.\ users said the platform was somewhat or very important for helping them understand what is happening in the world~\cite{pewresearchManyTurnYouTube2018}.
The role of the platform is also by no means limited to the United States~\cite{statista_youtube_2016}, attracting content creators that range from prolific music labels in India~\cite{chow_how_2018} to reactionary influencer--politicians in Brazil~\cite{fisher_how_2019}.

Given the sheer size of the platform's catalog and the multimedia nature of its content, systematically finding content on \YT is hard.
Thus, when researchers study the platform, they resort to a variety of heuristics such as 
using the website's search functionality~\cite{madathil_healthcare_2015}, 
snowball sampling recommendations~\cite{hortaribeiroAuditingRadicalizationPathways2019}, and 
selecting links that are shared on other platforms such as Twitter~\cite{wu_beyond_2018}.
These sampling strategies involve a great amount of extra work and often hinder the generalizability of studies (since the research is carried out with little knowledge about the representativeness of the data analyzed).

While the role of \YT in our lives is increasing, recent moves by platforms limiting access to data through application programming interfaces (APIs) exacerbate data collection challenges~\cite{bruns_after_2019}.
In this context, hoping to foster research about the platform, we present \textit{\datasetname,} a large-scale dataset of channel and video metadata from English-language \YT. 
To the best of our knowledge, this dataset represents the largest collection of \YT metadata made publicly available to date.

The dataset is composed of the following parts:
(1)~channel metadata for over \numchannelsapprox channels, including their numbers of subscribers and videos, as well as creation dates;
(2)~video metadata for over \numvideosapprox videos, including descriptions, numbers of likes, views, etc;
(3)~for most of the channels (97\%), time series of the numbers of subscribers and views at weekly granularity (2.8 years of data per channel on average);
(4)~channel-level weights to partially control for sampling biases; 
(5)~a table specifying for each of \numuserscomments anonymous users on which videos they commented.

\begin{figure*}[t]
    \centering
    \includegraphics[width=0.85\linewidth]{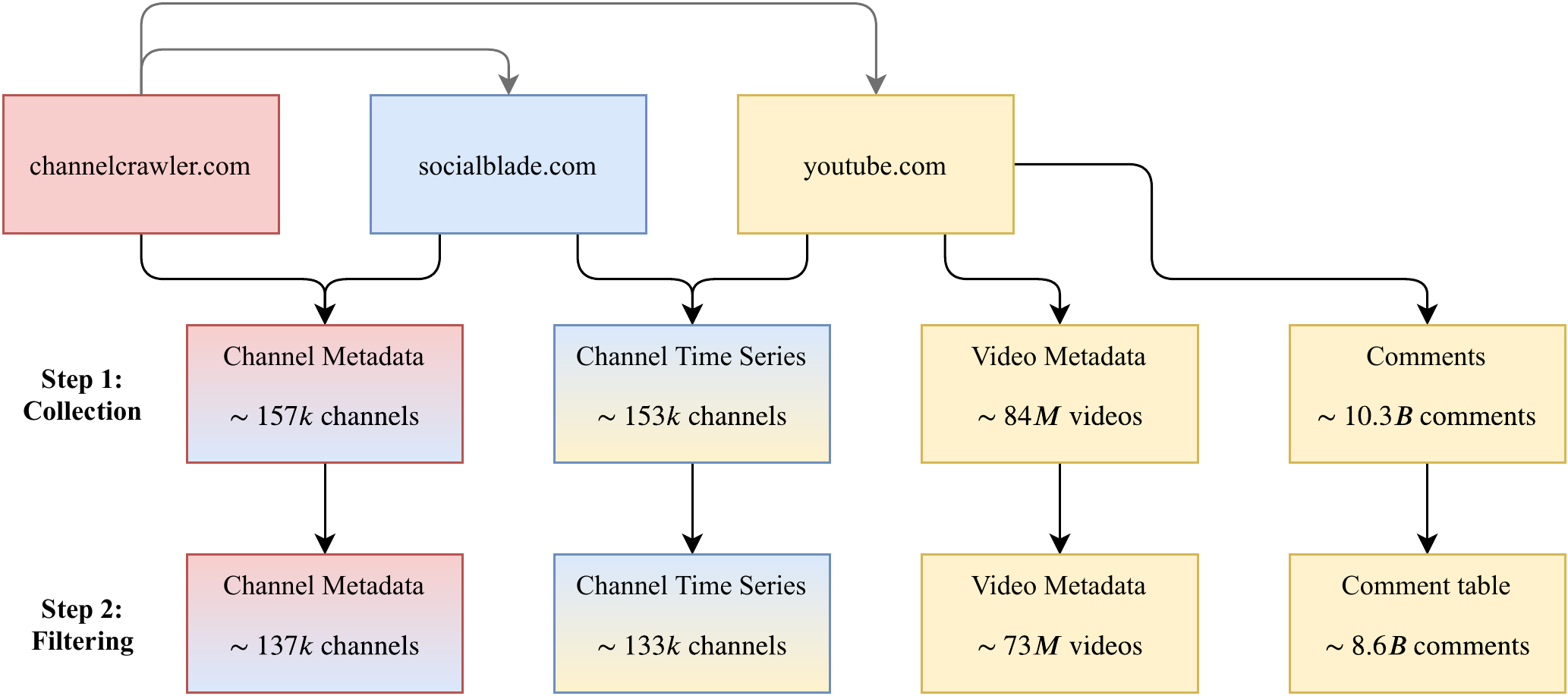}
    \caption{\textbf{Overview of our data collection and processing pipeline.} Leveraging three distinct data sources, we collect channel metadata, channel-related time series (i.e., the weekly number of viewers and subscribers), video-related metadata and comments. 
    We additionally filter this data using a standard language detection tool to ensure videos are from English-speaking channels. For comments, we release only a table specifying for anonymous users on which videos they commented (without the textual comments themselves). }
    \label{fig:yt_all}
\end{figure*}

\section{Dataset collection and preprocessing}
Besides \YT itself, we collected data from two third-party sources that aggregate \YT metadata:

\begin{enumerate}
\item \CC, a website that compiles \YT channels and makes them searchable through a variety of criteria.
\footnote{https://channelcrawler.com/eng/faq} 
The website has existed since mid-2013%
\footnote{\url{https://www.channelcrawler.com/}}
and uses a snowball sampling approach to collect channels.
\item \SB, a website that tracks social media statistics and analytics.
The website has existed since 2008 and originally tracked statistics for {digg.com}. 
In 2010, they switched their focus to tracking \YT statistics\footnote{https://socialblade.com/info} and have since added other platforms such as Twitch.tv and Instagram.
In 2014, the company launched consulting and channel management services to help content creators and companies that want to advertise on them.%
\footnote{https://socialblade.com/blog/nine-years-of-socialblade/}
\end{enumerate}

We illustrate the sources and the data collection and processing methodology in Fig. \ref{fig:yt_all}.

\subsection{Step 1: Collection}

\xhdr{Channel metadata.} 
We gathered a pool of \numchannelsraw channels by crawling all English channels with more than 10k subscribers and 10 videos from \CC. 
Data collection took place between 2019-09-12 and 2019-09-17.
Language classification was performed by the website using an automatic classifier.
Their repository is particularly helpful as \CC has been expanding its channel pool since 2013 (when scraping data from \YT was easier). 
We crawled additional metadata about \textit{subscriber rankings} for each of the channels from \SB, which order the channels according to their number of subscribers.
For example, if a channel's subscriber rank equals 10, this means that, among channels tracked by {\SB}, there are nine channels with a higher number of subscribers.
As discussed in Sec.~\ref{sec:007_ranking}, this allows us to assess the representativeness of our channel sample and possibly enhance it by weighting channels when performing data analysis.

\begin{figure*}[t]
    \centering
  \includegraphics[width=0.95\linewidth]{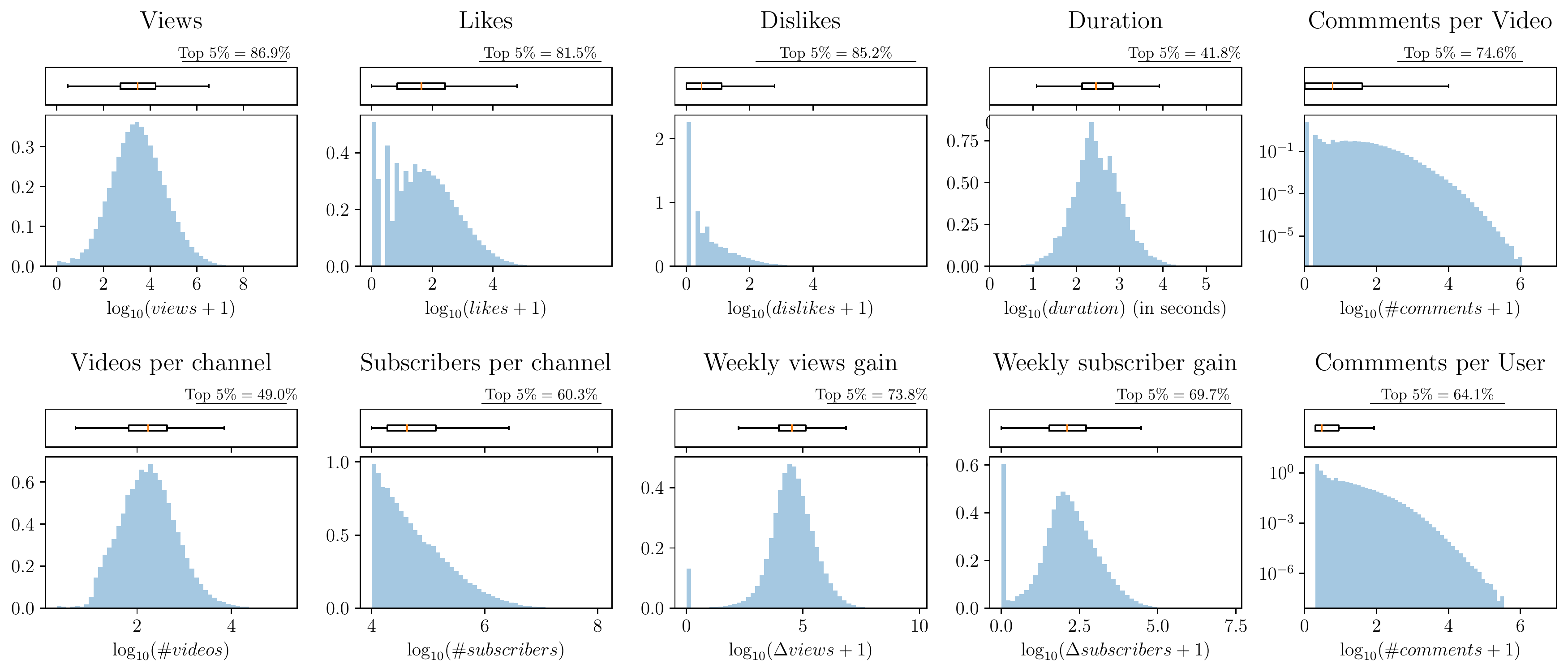}  
\caption{
\textbf{Distributions of video and channel statistics.}
Using both a box-plot and a histogram, we depict log-transformed distributions of video and channel statistics derived from video metadata (views, likes, dislikes, duration), channel metadata (videos per channel, subscribers per channel), time-series data (weekly views gain, weekly subscriber gain), and the comment table (comments per video, comments per user).
For the box-plot, box boundaries mark quartiles, the middle bar the median, and whiskers the 5th and 95th percentiles.
Above each plot, we report what percentage of the sum of each statistic is concentrated in the top 5\% of data points, as well as the range of values taken by the top 5\% (solid line). 
}
\label{fig:video_metadata}
\end{figure*}

\xhdr{Video metadata and comments.} 
For all \numchannelsraw channels obtained from \CC, we downloaded the metadata and comments for all their available videos from \YT.
In total, we crawled over \numvideosraw videos and \numcommentsraw comments between 2019-10-29 and 2019-11-23.

\xhdr{Channel time-series.}
Lastly, we compiled time series related to each channel. 
These come from a mix of \YT data and time series crawled from \SB.
From the former, we derived weekly time series indicating how many videos each channel had posted per week. 
From the latter, we crawled weekly statistics on the number of viewers and subscribers per channel. 
This data was available for around 153k channels. 

\subsection{Step 2: Filtering}
Although \CC automatically classifies the languages of each channel, we found that many of the channels labeled as English actually posted most of their content in Hindi or Russian. 
Hence, in order to ensure the consistency of the dataset, we additionally filtered channels using the \texttt{langdetect} library~\cite{shuyo2010language}. 
For each channel, we appended the titles and descriptions for 10 random videos and accepted those predicted to be in English with a probability above 60\%. 
This reduced the channel pool to \numchannelsexact and the video pool to \numvideosexact.

For comments, our filtering step was more aggressive. Besides removing all comments from non-English channels, we anonymized users by replacing their names with randomly assigned IDs and strip comments of their textual content.
We release only a table that specifies, for each anonymous user, which videos they have commented on.
To avoid the possibility of user de-anonymization, we only consider comments on videos with more than 30 commenters (as it would be easier to match an anonymized ID with a specific account if there are few comments).

\subsection{Ethics}
We collected only data publicly available on the Web and
did not (1)~interact with online users in any way nor (2)~simulate
any logged-in activity on \YT or the other platforms. 
Data was only collected for channels that had more than 10,000 subscribers and that had previously been collected and made searchable by third parties.
The comment table that we release does not allow for easy identification of users unless one crawls a substantial amount of YouTube. YouTube accounts are not typically linked with sensitive information.

\section{Dataset description}
The \datasetname dataset consists of four files, each of which we describe below.

\xhdr{Video metadata} (\texttt{\url{df_videos_en.jsonl.gz}}): 
Metadata associated with the \numvideosexact videos.
This includes videos from all \numchannelsexact channels.
The metadata include 
the category of channels (self-defined when they upload a video to  YouTube), channel and video IDs, upload date, number of likes, dislikes, and views, duration in seconds, textual description, and self-assigned tags.
Importantly, this data is obtained \textit{at crawl time} (between 2019-10-29 and 2019-11-23), which is also provided as a field. 

\xhdr{Channel metadata} (\texttt{\url{df_channels_en.tsv.gz}}): 
Metadata associated with the \numchannelsexact channels: channel ID, join date, country, number of subscribers, most frequent category, and the channel's position in \SB's subscriber ranking. 
The number of subscribers is provided both as obtained from \CC (between 2019-09-12 and 2019-09-17) and as crawled from \SB (2019-09-27).
Additionally, we also provide a set of weights (derived from \SB's subscriber rankings) that can be used to partially correct sample biases in our dataset (see Sec.~\ref{sec:007_ranking}).

\xhdr{Time-series data}
(\texttt{\url{df_timeseries_en.tsv.gz}}):
Time series of channel activity at weekly granularity.
The span of time series varies by channel depending on when \SB started tracking the channel. On average, it contains 2.8 years of data per channel for 133k channels (notice that this means there are roughly 4k channels for which there is no time-series data).
Each data point includes the number of views and subscribers obtained in the given week, as well as the number of videos posted by the channel.
The number of videos is calculated using the video upload dates in our video metadata, such that videos that were unavailable at crawl time are not accounted for.
We provide a brief characterization of the coverage of the time-series data in Sec.~\ref{sec:006_char}.

\xhdr{Comment table} (\texttt{\url{youtube_comments.tsv.gz}}):
A table that specifies for anonymized users on which videos they commented.
The table was constructed based on \numcomments comments made by \numuserscomments users on \numcommentsvideos videos. 
For each comment, we provide an anonymous user ID, a video ID, the number of replies the comment received, and the number of likes the comment received (again, these were obtained at crawl time, between 2019-09-12 and 2019-09-17).

\xhdr{Raw files.} 
For reproducibility purposes, we also make available the raw files, that is, the data files obtained from the websites before filtering out about 27k channels due to their non-English language. 
These raw files are made available in the same formats and with the same fields as the previously described files. 
Scripts to obtain the final files from raw files are also made available.

\section{Compliance with FAIR principles}

\datasetname conforms to the FAIR principles.
It is \textit{findable,} as it is made publicly available via Zenodo.
It is \textit{accessible,} as it can be accessed by anyone in the world and as it leverages standard data formats (\texttt{.tsv}, \texttt{.json}, \texttt{.gz}).
It is \textit{interoperable,} as almost every programming language has libraries that allow individuals to work with data in the formats employed.
And it is \textit{reusable,} as it is richly described in this paper.

\section{Dataset characterization}
\label{sec:006_char}

Next, we provide a quantitative overview of the dataset, describing the data and assessing its completeness. 

\xhdr{Video and channel statistics.} 
Fig.~\ref{fig:video_metadata} presents log\hyp transformed distributions associated with different video and channel statistics.
In the first four columns of the top row, we show statistics derived from video-level metadata.
Using a histogram and a box-plot, we depict the distributions of the number of views, likes, dislikes, and video duration.
Note that the video duration distribution is bimodal, with peaks around $2^{2.25}$ (about 3 minutes) and  $2^{2.75}$ (about 10 minutes).
The second peak may be explained by a threshold set by YouTube, where content creators could place multiple ads if a video was longer than 10 minutes~\cite{alexander_youtube_2020}.\footnote{YouTube has since changed the threshold to 8 minutes: \url{https://support.google.com/youtube/answer/6175006}}

In the first four columns of the bottom row, we depict both channel-level statistics (in the first two columns, from \CC data), and time series--related statistics (in the last two columns, from \SB data).
We note that there is a high number of weeks for which the weekly gain of views and subscribers equals zero (about 2\% of view data points and about 8\% of subscriber data points). 
While these may be due to data collection or corrections done by \SB, we find that most of the missing data points for subscriber data are associated with channels with close to 10,000 subscribers. 
More specifically, around 76\% of missing data points belong to channels that had between 9,900 and 10,100 users. 
We believe this could be indicative of YouTubers artificially boosting their channels towards the 10,000\hyp subscriber mark (and subsequently not gaining any additional subscribers).

Finally, in the last column, we depict the distribution of comments per video and comments per user in the first and second row, respectively. Note that, unlike the other plots, here we also plot the $y$-axis on a logarithmic scale, since the distribution is heavy-tailed. 

\begin{figure}
    \centering
    
    \includegraphics[width=\linewidth]{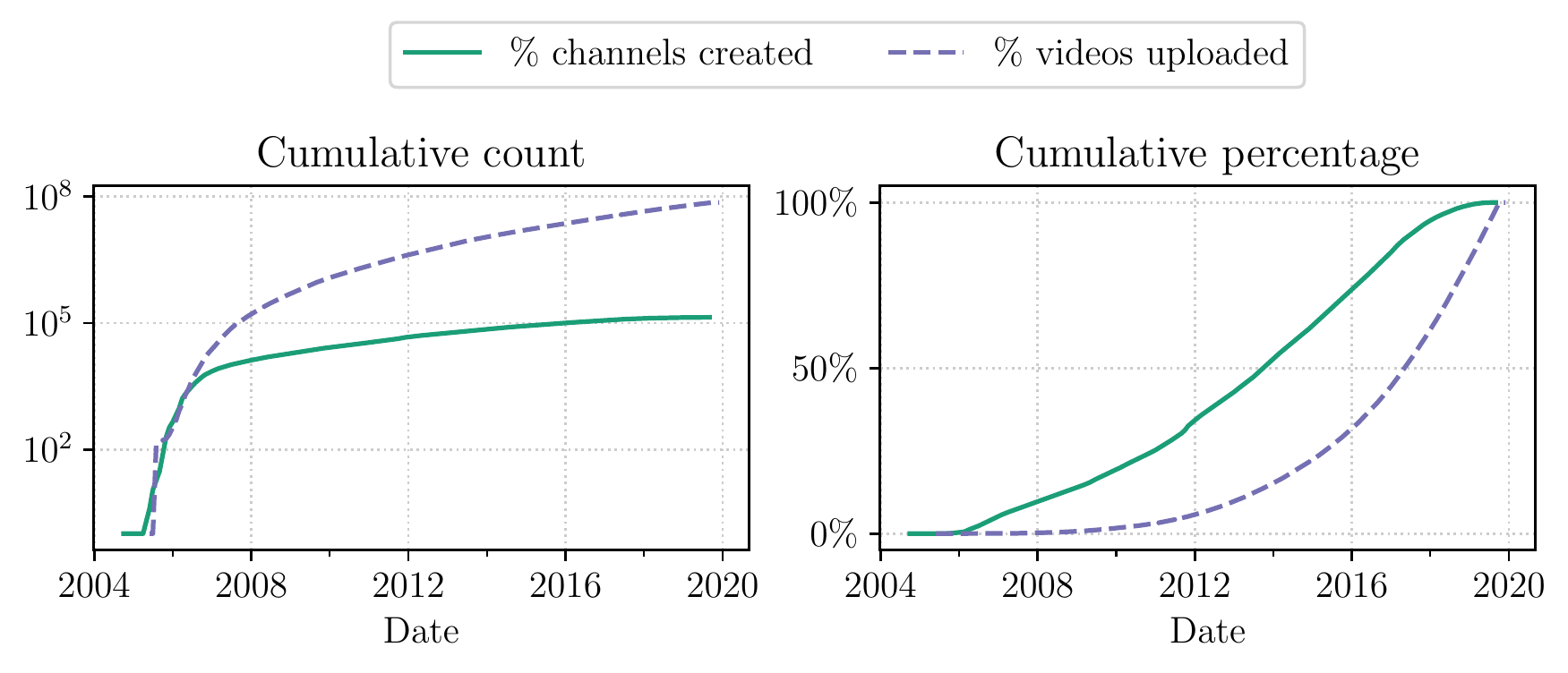}
    \caption{\textbf{Temporal distribution of video and channel creation.} Cumulative number of videos and channels created at each point in time from 2004 to 2019 (left); normalized by the total number of videos and channels (right).}
    \label{fig:overall_ch}
    
    \includegraphics[width=\linewidth]{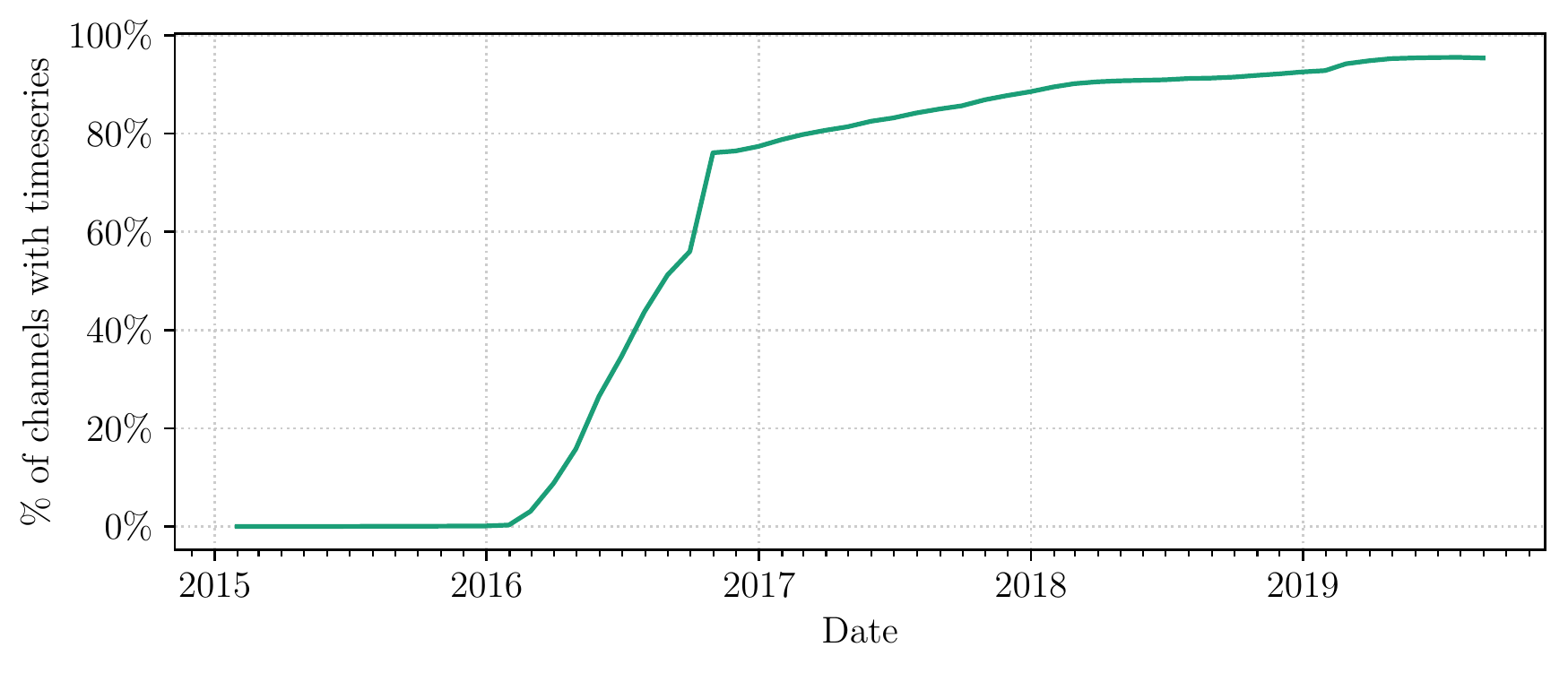}
    \caption{\textbf{Time series completeness.} Monthly percentage of channels for which there is time-series data.
    For each value on the $x$-axis, we plot the percentage of channels that were already created for which time-series data is available.}
    \label{fig:complet_ts}

    \includegraphics[width=\linewidth]{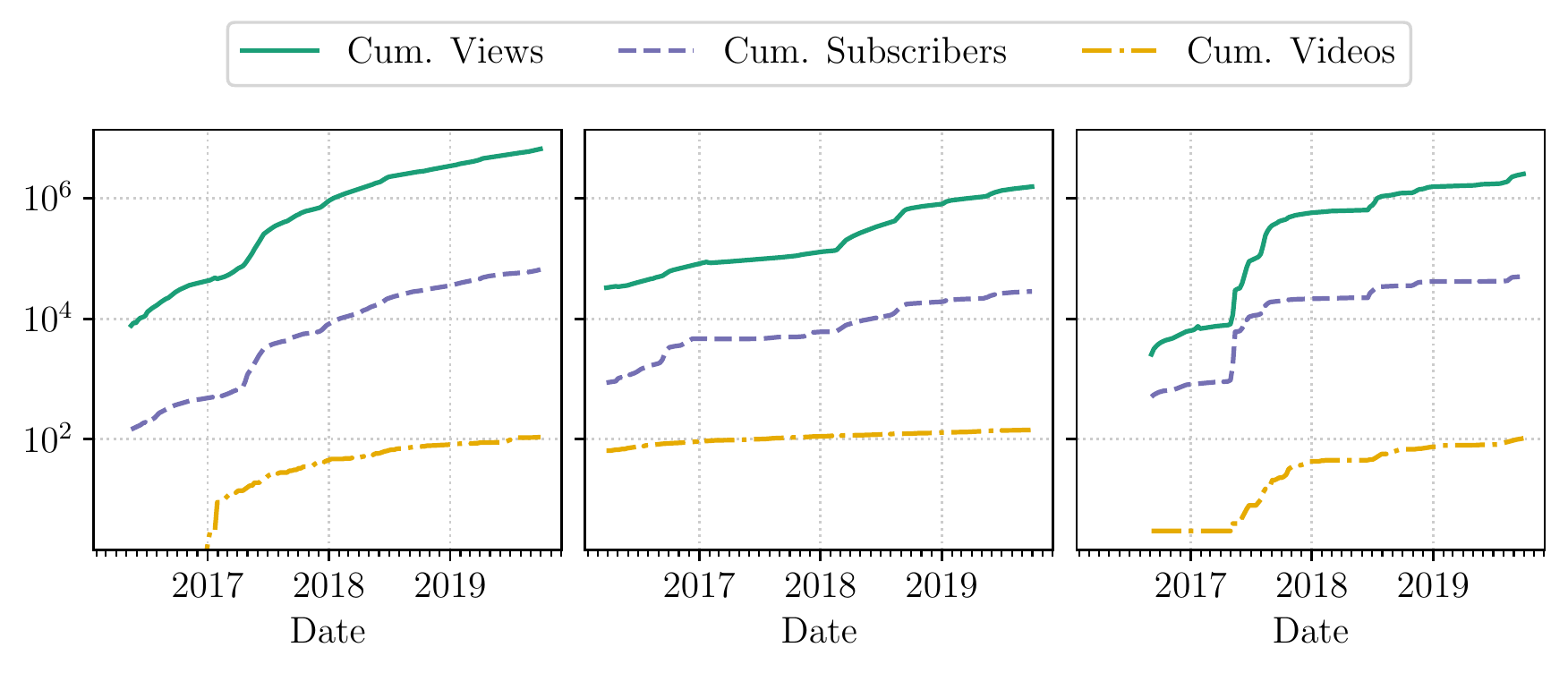}
    \caption{\textbf{Time-series examples.} For three channels selected at random,
    we depict the cumulative number of views, subscribers, and videos. }
    \label{fig:example_ts}

    \includegraphics[width=\linewidth]{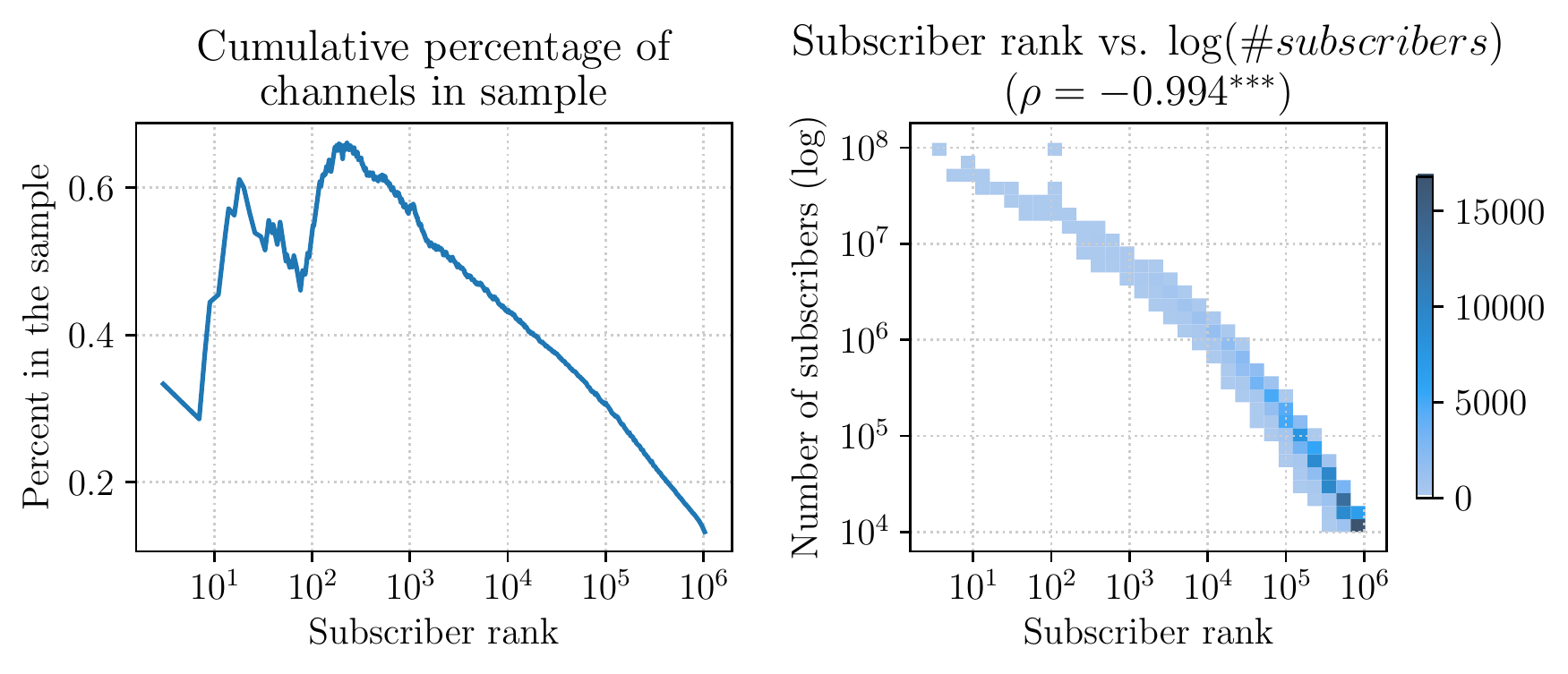}
    \caption{
    \textbf{Ranking completeness.}
    On the left-hand side, we depict, for each subscriber rank (obtained from \SB) in the dataset, the percentage of channels present in the dataset up to that rank. 
    On the right-hand side, we depict the relationship between subscriber rank and number of subscribers (obtained from \CC).
    }
    \label{fig:ranking}
\end{figure}

On top of the box-plot associated with each statistic, we draw a line showing the range of values taken by the top 5\% of the data points.
Above the line, we also report what percentage of all values belongs to the top 5\%.
We find that some of the distributions are highly skewed.
For example, in accordance with previous work~\cite{rieder_mapping_2020, bartl_youtube_2018}, the videos with the top 5\% most views are responsible for 86.9\% of all views and the top 5\% most commented videos are responsible for 74.6\% of all comments.

\begin{table}[t]
\centering
\footnotesize
\caption{Total number of views, videos, and likes, and total video duration in each category.}
\label{tab:sumv}
\begin{tabular}{lrrrr}
\toprule
Categories &  Views &  Videos &  Likes &  Duration \\
  &  ($\times 10^9$) &   ($\times 10^6$)  & ($\times 10^6$) &  (years) \\
\midrule
Autos \& Vehic.      &        123.1 &           2.3 &        956.3 &              27.0 \\
Comedy                &        345.0 &           1.2 &       6134.2 &              18.9 \\
Education             &        502.4 &           3.8 &       2881.4 &             117.1 \\
Entertainment         &       2287.7 &          12.3 &      20140.6 &             243.1 \\
Film \& Anim.      &        576.3 &           2.4 &       3737.8 &              47.7 \\
Gaming                &       1042.5 &          13.7 &      17351.3 &             623.2 \\
Howto \& Style         &        420.7 &           4.0 &       5790.8 &              73.6 \\
Music                 &       2475.0 &           8.3 &      19270.5 &             124.1 \\
News \& Pol.       &        158.6 &           8.9 &       1892.9 &             144.8 \\
Nonprofits &         18.5 &           0.8 &        242.9 &              29.5 \\
People \& Blogs        &        617.3 &           6.9 &       9138.1 &             148.8 \\
Pets \& Animals        &         72.1 &           0.6 &        659.8 &               9.9 \\
Science \& Tech.  &        175.0 &           2.4 &       1978.4 &              47.5 \\
Sports                &        262.0 &           4.4 &       2528.1 &              77.3 \\
Travel \& Events       &         56.4 &           1.1 &        459.8 &              19.2 \\ \midrule
Total                 &       9132.7 &          72.9 &      93162.9 &            1751.8 \\ \bottomrule
\end{tabular}

\end{table}

\xhdr{Aggregated video statistics.} 
Table~\ref{tab:sumv} shows the total number of views, videos, and likes, as well as the total duration for each of the 15 categories provided by the platform. 
We find that \textit{Music} and \textit{Entertainment} are the most popular categories (with over 2~trillion views each), while the \textit{Gaming} category contains the most videos (over 13.7M videos).
\textit{Gaming} videos are also substantially longer than those in other categories: while the 12.3M videos in the \textit{Entertainment} category amount to around 243 years, the 13.7M \textit{Gaming} videos amount to more than 600 years.

\xhdr{Video and channel creation dates.} Fig.~\ref{fig:overall_ch} shows the creation dates of videos and channels in the dataset in both relative and absolute terms.
We find that around 50\% of channels in the dataset were created after 2014, and around 50\% of videos were created since 2018. 
The fact that the median creation date for videos is more recent than that for channels could be due to \YT's growth over the years.
However, it is worth noting that older channels and videos may have been deleted by content creators or taken down following copyright complaints, which could induce a recency bias in our video sample.

\xhdr{Time-series completeness.} Notice that not all channels have the same time series data available. We characterize the completeness of our data over the years in Fig.~\ref{fig:complet_ts}. For each date, shown on the $x$-axis, we plot the percentage of channels that were created by that point for which time-series data is available.
Starting in late 2016, most of the channels (about 80\%) have time series data available.
We additionally illustrate the time-series data made available with three randomly sampled channels in Fig.~\ref{fig:example_ts}, depicting the cumulative number of views, subscribers, and videos of each channel throughout their full lifespan.

\xhdr{Estimating representativeness via subscriber ranks.} 
In Fig.~\ref{fig:ranking} (left), we explore \SB rankings to get a sense of how representative our data is compared to \SB's catalog
We find that for the top 10k channels, we have around 35\% of all channels present in \SB subscriber ranking, and for the top 100k, around 25\%.
As a sanity check, we additionally study the relationship between the subscriber ranks provided by \SB and the actual number of subscribers of each channel (obtained from \CC). 
Although there are discrepancies between the two, we observe a very high Spearman correlation coefficient  ($-0.99$).

\section{Correcting for sampling bias}
\label{sec:007_ranking}
\begin{figure}[t]
    \centering
    \includegraphics[width=0.8\linewidth]{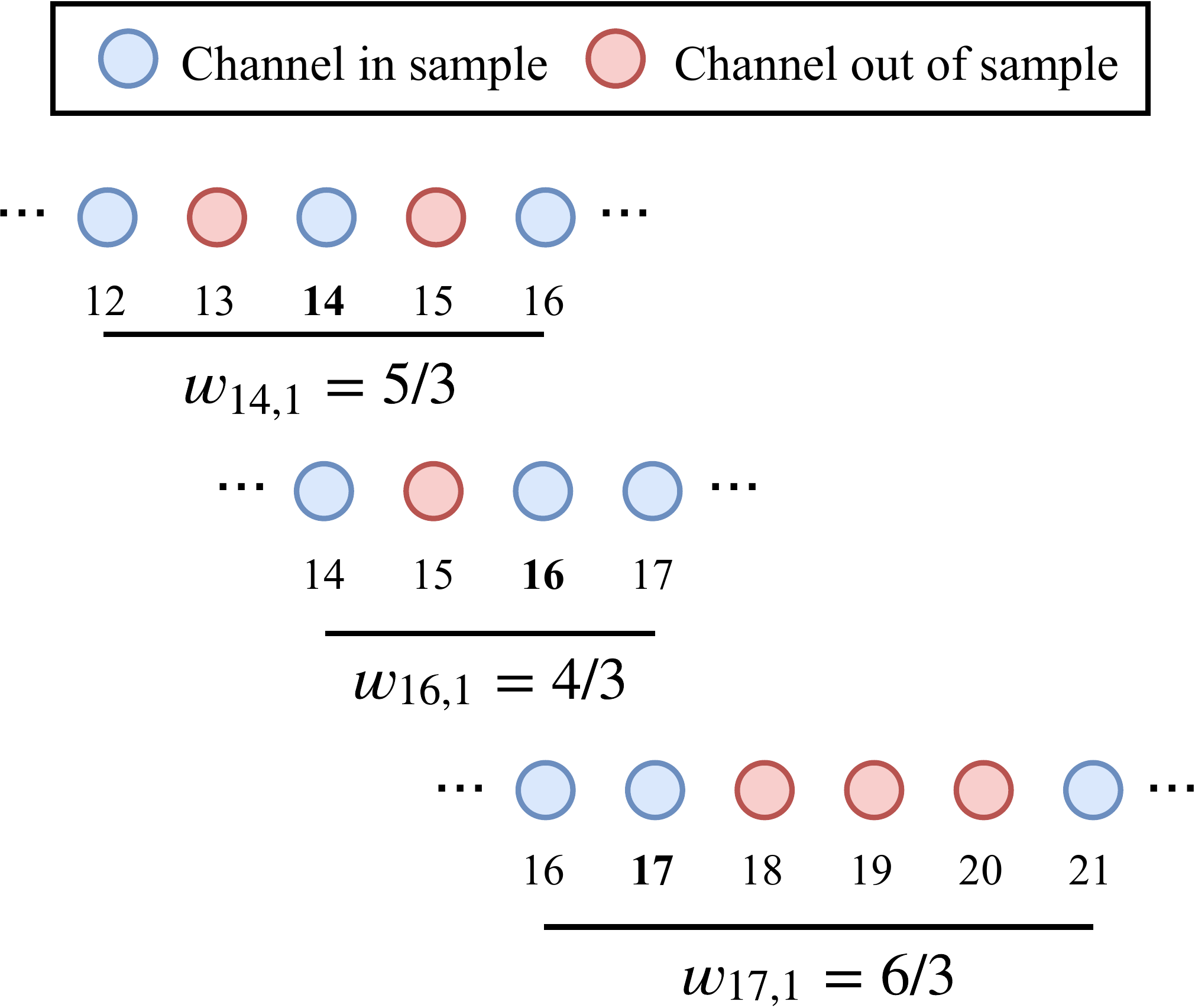}
    \caption{\textbf{Weighting scheme.} 
    Suppose that we are assigning weights to three channels present in the \datasetname sample (14, 16, 17) using a window size $k = 3 = 2m+1$. 
    This means that when assigning a channel its weight, we look for the first ($m = 1$) channel  in the sample before and after the channel at hand. 
    So, e.g., if we have channels 12 and 16 in the sample, but not 13 or 15, we would use a weight of $5/3$ for channel 14, i.e., the inverse percentage of channels sampled in the window around channel 14: $w_{14,1} = P(i=14;m=1)^{-1} = ((2\cdot 1+1)/(16-12+1))^{-1} = 5/3$. 
    If the window around a channel spans a narrower range of ranks (e.g., for 16) the weight will be smaller; if the window spans a wider range of ranks (e.g., for 17) the weight will be larger. }
    \label{fig:weights}
\end{figure}

\begin{figure}[t]
    \centering
    \includegraphics[width=\linewidth]{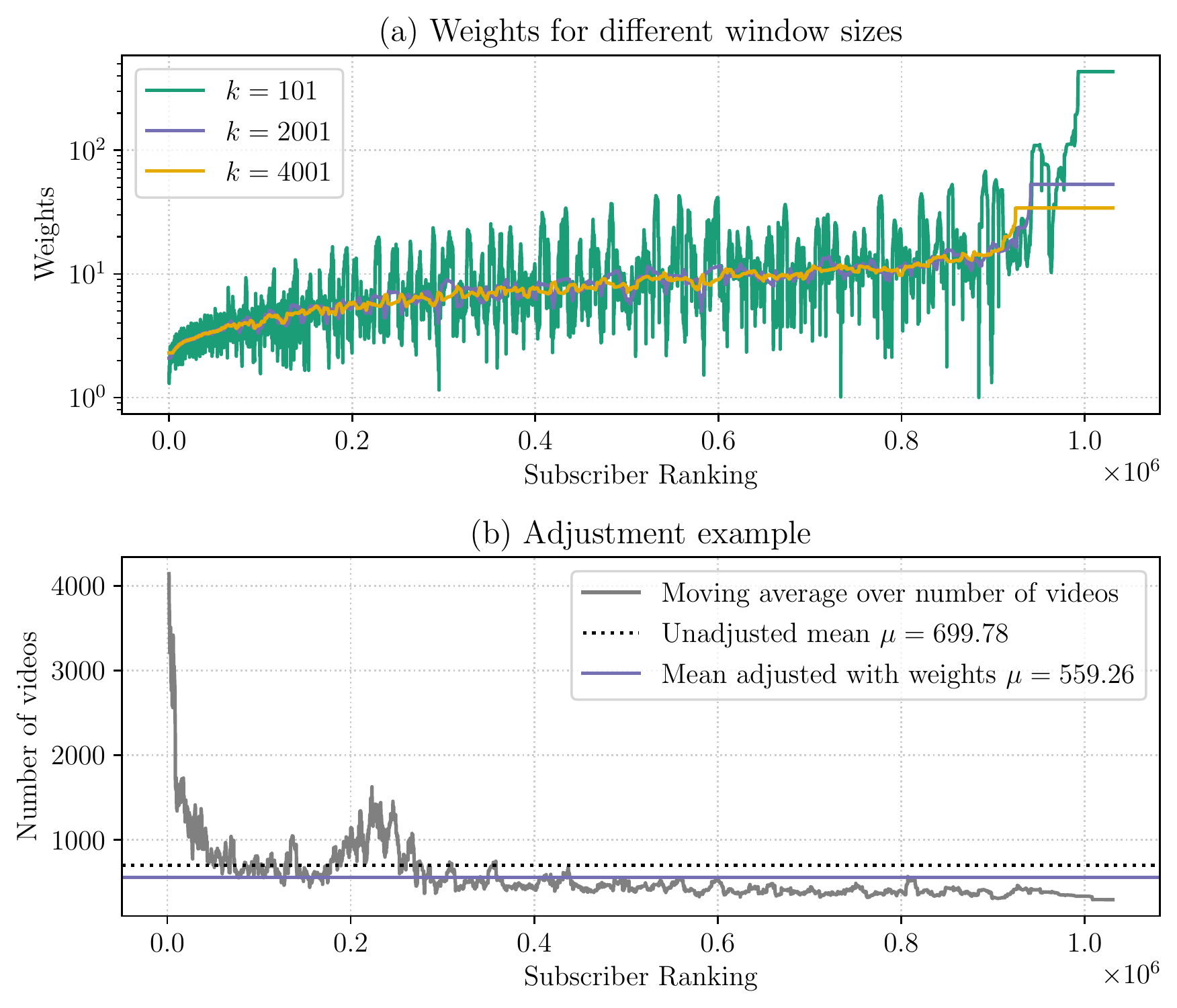}
    \caption{\textbf{Weights in practice.} 
    In \textbf{(a)}, we depict how our method assigns weights to channels of different subscriber ranks. 
    In \textbf{(b)}, we depict a moving average over the ranks (window size 1000) that calculates, at each step, the average number of videos per channel.
    Two horizontal lines depict the average value with and without the weight adjustment.}
    \label{fig:weights_window}
\end{figure}

Fig.~\ref{fig:ranking} shows that our sample is biased towards channels higher up in \SB's ranking with respect to the number of subscribers.
In this section, we discuss how we can account for this bias in subsequent analyses by giving more weight to channels from rank ranges that are under-represented in \datasetname.
Note that \SB keeps track of over 23 million YouTube channels\footnote{https://socialblade.com/info} through the official YouTube Data API, which suggests that their catalog comprises an expressive share of relevant YouTube channels. Unfortunately, we do not know how representative Social Blade's channel repository is of YouTube, but the authoritativeness with which the YouTube community leverages the platform suggests that it contains a good share of the relevant channels on the platform.

Let $L$ be an array that contains \SB ranks for the channels included in the \datasetname dataset, sorted by subscriber counts in decreasing order;
i.e., $L[i]$ is the rank (among \textit{all} of \YT's channel, according to \SB) of the channel with $i$-th largest number of subscribers among all channels contained in \datasetname.
For the $i$-th-most followed channel in \datasetname and an odd-valued window of size $k = 2m + 1$, its \emph{local sampling probability} is
$$
P(i; m) = \frac{2m + 1}{L[i + m] - L[i - m] +1}.
$$
$P(i;m)$ is the percentage of all \YT channels between ranks $L[i + m]$ and $L[i - m]$ that are contained in \datasetname.
Thus, in order to correct for the sampling bias, we may weight each channel by its inverse local sampling probability, i.e., by
$$
w_{i,m} = P(i; m)^{-1}.
$$

We illustrate this procedure in Fig.~\ref{fig:weights}, but the rationale is simple: while the probability of sampling channels of different ranks is clearly not homogeneous globally, we can estimate the \textit{local} probabilities and assign to channels weights inverse to those probabilities. 
In our described method, the size $k$ of the window controls in how local a manner we estimate the sampling probability.%
\footnote{Notice that another way to obtain local sampling probabilities would be to use a window of fixed size and slide it through the rankings, calculating the sampling probability at different points. 
Empirically, we found that the method proposed yielded less noisy results and was capable of better handling the sparse nature of our data at lower ranks. 
Nevertheless, different weighting schemes could be implemented using the data at hand and the code provided.
}

\xhdr{Choosing the window size \textit{k}.}
An unsolved issue with the proposed method is how to determine the window size $k$.
An overly small window size may create noisy weights, while an overly large window size may defeat the purpose of the method altogether (after all, the whole point is to control for \emph{local} differences in sampling probability).
To empirically determine a good window size, we experiment with a wide range of values\footnote{$k \in \{101, 251, 501, 1001, 2001, 4001, 8001, 16001, 32001\}$} and pick the smallest value for which we find the time series to appear smooth ($k=2001$).
We illustrate the weights for three different values of $k$ in Fig.~\ref{fig:weights_window}(a). 

\xhdr{Example usage.}
We provide a concrete example of how to use these weights to obtain more representative statistics.
Suppose, for instance, that we want to estimate the average number $\mu$ of videos per channel on YouTube. 
If we na\"ively take the average of our sampled channels, we find that $\mu=699.78$. 
However, notice that channels in the top ranks have on average many more videos, as shown in Fig.~\ref{fig:weights_window}(a). 
Thus, since these channels are much more likely to be in the sample, we are over-estimating the average number of channels.
Fortunately, we can correct for our bias by calculating the \emph{weighted} average, finding $\mu=559.26$.
Interestingly, we find little variation from the different window sizes in this estimate, for example, for $k=101$, we find $\mu=555.53$, and for $k=32001$, we find $578.09$.

\xhdr{Limitation of the correction scheme.}
It is important to note that our dataset focuses on English-language YouTube channels, while the subscriber rank is language\hyp agnostic.
This implies that we could use this sample to deduce information about \YT in general.
However, when doing so, we are likely to over-emphasize features associated with English\hyp language channels.
For example, if we try to estimate the average number of videos per channel in all of YouTube using only our sample, and if it happens that the average number of videos per channel for English channels is smaller than for other languages, we may under\hyp estimate the statistic.
Moreover, our weighting methodology presupposes that \SB rankings are complete and correct.
Regardless of these limitations, we argue that this method is far superior to what has been done in previous work (which we extensively review in the next section), where the representativeness of the data is often neither considered nor assessed.

\section{Related work}

We review previous studies using YouTube data with an emphasis on their data collection efforts. 
We emphasize work that has made data available and that tried to broadly characterize YouTube quantitatively.


\xhdr{Performance-oriented characterizations.} 
Analyses that characterized YouTube were conducted as early as 2007. 
Back then, the main interest was to characterize content creation and consumption patterns on the platform, often with a focus on enhancing the service quality in terms of performance~\cite{gill_youtube_2007,gill_characterizing_2008,zink_characteristics_2009}.
In this direction, we highlight the works of \citet{cha_i_2007} and of \citet{cheng_statistics_2008}, which characterized YouTube in terms of how much content was produced and consumed, comparing statistics with other video-on-demand systems.
In both cases, the authors identified patterns in the creation and consumption of videos that could be leveraged to enhance service quality (which was key to scale video-on-demand systems).

The two papers employ different data collection methodologies.
\citet{cha_i_2007} crawled and analyzed the \textit{Entertainment} and \textit{Science and Technology} categories, which had nearly 2~million videos altogether.
When data for the paper was collected, YouTube had pages indexing all videos that belonged to a given category,  which made data collection viable.
\citet{cheng_statistics_2008} performed a four-month crawl collecting metadata for over 3~million YouTube videos, using a breadth-first search to find videos via recommendations.

As the platform grew and challenges related to its infrastructure increased, performance-oriented analyses of the website continued to be an active area of research~\cite{finamore_youtube_2011, orsolic_machine_2017, schwind_dissecting_2020}. 
Of this subsequent work, of particular relevance is the method developed by \citet{zhou_counting_2011} to obtain an unbiased sample of YouTube videos via random prefixes. 
They leveraged a feature of YouTube's search API where one could match the prefix of video IDs by using queries such as \texttt{watch?v=xyz*}.
Their analysis estimated that there were around 500 million YouTube videos at the time of the research, and shed light on the bounds on the total storage YouTube must have had, as well as the network capacity needed to deliver videos.

\xhdr{Virality and engagement.} 
Previous research has also leveraged YouTube data to characterize the dynamics of virality, as well as to better understand which factors (content-related or not) make videos popular.

\citet{figueiredo_tube_2011} characterized the popularity growth of \YT videos over time.
Leveraging a now\hyp deprecated feature that allowed to extract time-series data for statistics such as ratings and views, they collected data associated with over 150k unique videos from \YT top lists, from the YouTomb project%
\footnote{An initiative to monitor videos removed due to copyright violations.}, and by using the search functionality with random topics obtained from a lexical ontology. 
They found popularity growth patterns to be largely dataset-reliant. For example, while videos in the top lists experienced sudden bursts of popularity throughout their lifetime, copyright\hyp protected videos received most of their views earlier in their lifetime.

In a similar vein, \citet{borghol_characterizing_2011} studied the popularity dynamics of user-generated videos leveraging data obtained by sampling recently uploaded videos and videos obtained through keyword search, finding significant differences from results between the two samples.
In subsequent work~\cite{borghol_untold_2012}, they crawled a large dataset of 48 sets of identical or nearly identical videos (1,761 videos overall) to study content-agnostic factors that impacted YouTube video popularity. They found a strong ``rich-get-richer'' behavior, with the total number of previous views being the most important factor for predicting which videos would gain views.

\citet{brodersen_youtube_2012} investigated the relationship between popularity and locality of online YouTube videos, finding that, despite the global nature of the Web, video consumption is largely constrained by geographic locality. The paper, authored by Google employees, leveraged a random sample of 20M YouTube videos uploaded between September 2010 and August 2011.

\citet{abisheva_who_2014} analyzed how YouTube videos were shared on Twitter. 
Their dataset comprised 5.6 million YouTube videos and over 15M video-sharing events from around 87k Twitter users.
Among other things, their findings suggest a super-linear relationship between initial video success on Twitter and final success on YouTube.

\citet{wu_beyond_2018} studied whether engagement in online videos could be predicted. 
The authors defined a new metric, \textit{relative engagement,} which they found to be strongly correlated with recognized notions of quality. 
To obtain the data for this study, the authors developed a crawler to collect three daily time series related to video attention dynamics: the volume of shares, view counts, and watch time (leveraging the same feature as \citet{figueiredo_tube_2011}). 
They analyzed two datasets: a collection of over 5.3M videos published between 2016-07-01 and  2016-08-31 from around 1M channels collected from Twitter's streaming API, as well as over 96k videos from high-quality sources.

\begin{table*}[t]
\small
\centering
\begin{tabular}{p{3.7cm}p{3cm}p{3cm}p{5cm}}
                      & \textbf{Kind of data}   & \textbf{Dataset size} & \textbf{Sampling strategy} \\ \midrule
\citet{cha_i_2007} & Video metadata  & $\sim$2M videos & Crawled YouTube categories ``Entertainment'' and ``Science and Technology.'' \\ \midrule
\citet{cheng_statistics_2008}      & Video metadata  \newline (repeated crawls) & $\sim$3M videos & Multiple BFS crawls done between March 5th and April 16th 2007. \\ \midrule
\citet{borghol_characterizing_2011}   & Video metadata \newline (repeated crawls) & $\sim$1.2M  videos & Recently uploaded videos +   \newline keyword search \\\midrule
\citet{abisheva_who_2014}    & Video metadata + \newline video sharing events  & $\sim$5.6M videos & Extracted videos from 28h of all public tweets containing URLs  \\\midrule
\citet{wu_beyond_2018} &Video metadata \newline popularity time-series &  $\sim$5.3M videos & Twitter Stream API (between July 1st and August 31st + high quality sources \\\midrule
\citet{bartl_youtube_2018} & Video metadata & $\sim$8M videos & Keyword search \\ \midrule
 \textbf{\datasetname} & Video metadata  + \newline
channel metadata + \newline comment data  + \newline popularity time-series & $\sim$\numvideosapprox videos   & Crawled websites that publicly display statistics about YouTube. \\\bottomrule                                        
\end{tabular}
\caption{Comparison between \datasetname and other \YT datasets made available in previous work. Notice that, for brevity's sake, we report the dataset sizes only in terms of the number of videos.}
\label{tab:datasets}
\end{table*}

\xhdr{Platform features.} 
Several other studies collected YouTube data to understand the impact or usage of specific features, such as comments, ads, or recommendations.

\citet{benevenuto_video_2009} studied ``response videos'', a now\hyp deprecated system where anyone could respond to a YouTube video with a video of their own. 
The response video would then be shown along  with the video to which it responded.
Analyzing around 196k YouTube users,  224k videos with responses, and 418k response videos, they found evidence of opportunistic behavior such as self-promotion and spamming.

\citet{zhou_impact_2010} studied the impact of the recommender system on video views, finding that recommendations are the main source of views for the majority of the videos on YouTube.
Leveraging the same deprecated feature as \citet{figueiredo_tube_2011}, they analyzed around 700k videos collected (1)~via the API or (2)~by capturing and parsing video requests at a university network gateway.

\citet{siersdorfer_how_2010} presented an in-depth study of commenting on YouTube. They analyzed the relationship between comments and views and trained a classifier capable of predicting the community's acceptance of a given comment. 
The authors analyze 6M comments on 67k YouTube videos obtained through the website's search functionality.

\citet{arantes_understanding_2016} leveraged logs of HTTP requests from a large university to study ad consumption on YouTube. 
Their dataset comprised around 99k video-ad exhibitions, 5.6k unique ads, and 58k unique videos.
Their analysis found the fraction of ad exhibitions that are streamed until completion to be high (around 29\%) relative to traditional online advertisements (where click-through rates are below 0.01\%).

\xhdr{Further characterizations.} 
Recent studies have again tried to provide overall characterizations of YouTube. 
We highlight two recent papers in this direction.
\citet{bartl_youtube_2018} obtained a sample of around 8M videos belonging to approximately 20k channels by randomly searching for keywords.
\citet{rieder_mapping_2020} performed what is perhaps the largest study characterizing YouTube, analyzing the static metadata of over 36M channels and 700M videos.
Both these studies provide high-level statistical analysis, finding, for example, that the vast majority of views goes to a small minority of channels.

\xhdr{Other studies.} 
So far, we have discussed attempts to broadly characterize YouTube, its features, and the dynamics of virality on the platform.
These are the most relevant previous works to contextualize the contribution provided by our dataset.
However, it is worth stressing that previous work has also explored problematic phenomena on the platform~\cite{sureka_mining_2010,ottoni_analyzing_2018,papadamou_disturbed_2019}, the quality of health-related information available~\cite{madathil_healthcare_2015,freeman_is_2007}, and the utility of the platform for online learning~\cite{alwehaibi_impact_2015,clifton_can_2011}.

\section{Discussion and conclusion}

In this last section, we briefly compare our dataset with existing large-scale YouTube\hyp related data that is publicly available.
Additionally, we discuss some research directions where we think \datasetname may be particularly useful.

\xhdr{Relationship between this and prior work.} 
We briefly discuss the relationship between \datasetname and the data previously used by researchers to better understand YouTube. 
We focus our comparison with other datasets that were made publicly available in previous work and depicted in Table~\ref{tab:datasets}.
Compared to previously available data, \datasetname is

\begin{itemize}
    \item \textbf{Channel-driven.} 
    For each channel, we collect  metadata and popularity time series along with all available videos.
    Other recent datasets focus largely on obtaining representative video samples~\cite{bartl_youtube_2018, wu_beyond_2018}. 
    Our approach is particularly interesting to study the process of content creation on YouTube (since our dataset has, for each channel, all videos available at crawl time).

    \item \textbf{Large.} 
    \datasetname is an order of magnitude larger than other recent publicly available YouTube dataset~\cite{bartl_youtube_2018}.
    This is particularly important due to the heterogeneity of YouTube. 

    \item \textbf{Recent.} 
    Unlike most previous comprehensive large-scale datasets\cite{cha_i_2007, cheng_statistics_2008}, \datasetname contains data from more recent years, where problematic phenomena on YouTube, such as troublesome children content~\cite{papadamou_disturbed_2019} or fringe content~\cite{hortaribeiroAuditingRadicalizationPathways2019}, moved into the spotlight.
    We hope that \datasetname enables the better contextualization of such content in the broader YouTube context.
\end{itemize}

\xhdr{Possible use cases.}
We believe this dataset can foster research on YouTube in a variety of ways.
First, it may help researchers to quantitatively study the evolution of \textit{content creation} on the platform throughout the years.
As ``digital influencers'' become increasingly important in the public debate, studying the way in which channels grow and professionalize is key to better understand our current information ecosystem.
Second, the dataset may help to study the evolution of content itself on YouTube.
Since its creation, the ``rules of the game'' have changed several times on the platform, and video metadata enables us to capture many of these transformations (e.g., what are the ideal video lengths throughout the years?).
Lastly, \datasetname can be a useful resource for a variety of more focused studies. 
As discussed, previous work often resorts to simple heuristics to find channels or videos related to specific topics (e.g., cancer), and \datasetname may act as a comprehensive starting point.

\section*{Acknowledgements}

We would like to thank Richard Patel for useful insights in all things related to crawling.
We also like to thank Timot\'e Vaucher and Jonathan Kaeser for making an awesome data visualization of \datasetname.
Our work is supported in part
by Swiss National Science Foundation grant 200021\_185043
and by gifts from
Google, Facebook, and Microsoft.

{
\small
\bibliography{refs}

\begin{thebibliography}{36}
\providecommand{\natexlab}[1]{#1}
\providecommand{\url}[1]{\texttt{#1}}
\providecommand{\urlprefix}{URL }
\expandafter\ifx\csname urlstyle\endcsname\relax
  \providecommand{\doi}[1]{doi:\discretionary{}{}{}#1}\else
  \providecommand{\doi}{doi:\discretionary{}{}{}\begingroup
  \urlstyle{rm}\Url}\fi

\bibitem[{Abisheva et~al.(2014)Abisheva, Garimella, Garcia, and
  Weber}]{abisheva_who_2014}
Abisheva, A.; Garimella, V. R.~K.; Garcia, D.; and Weber, I. 2014.
\newblock Who watches (and shares) what on Youtube? and when? Using twitter to
  understand Youtube viewership.
\newblock In \emph{WSDM}.

\bibitem[{Alexander(2020)}]{alexander_youtube_2020}
Alexander, J. 2020.
\newblock {YouTube} is finally letting creators know exactly how they’re
  making money on {YouTube}.
\newblock
  \url{https://www.theverge.com/2020/7/10/21319938/youtube-monetization-metric-ad-revenue-cpm-rpm-alternative-memberships-premium}.
\newblock Accessed: 2021-04-06.

\bibitem[{Alwehaibi(2015)}]{alwehaibi_impact_2015}
Alwehaibi, H.~O. 2015.
\newblock The {Impact} {Of} {Using} {YouTube} {In} {EFL} {Classroom} {On}
  {Enhancing} {EFL} {Students}' {Content} {Learning}.
\newblock In \emph{Journal of College Teaching \& Learning (TLC)}.

\bibitem[{Arantes, Figueiredo, and Almeida(2016)}]{arantes_understanding_2016}
Arantes, M.; Figueiredo, F.; and Almeida, J.~M. 2016.
\newblock Understanding video-ad consumption on {YouTube}: a measurement study
  on user behavior, popularity, and content properties.
\newblock In \emph{ACM WebScience}.

\bibitem[{Benevenuto et~al.(2009)Benevenuto, Rodrigues, Almeida, Almeida, and
  Ross}]{benevenuto_video_2009}
Benevenuto, F.; Rodrigues, T.; Almeida, V.; Almeida, J.; and Ross, K. 2009.
\newblock Video interactions in online video social networks.
\newblock In \emph{ACM Transactions on Multimedia Computing, Communications,
  and Applications}.

\bibitem[{Borghol et~al.(2012)Borghol, Ardon, Carlsson, Eager, and
  Mahanti}]{borghol_untold_2012}
Borghol, Y.; Ardon, S.; Carlsson, N.; Eager, D.; and Mahanti, A. 2012.
\newblock The untold story of the clones: content-agnostic factors that impact
  {YouTube} video popularity.
\newblock In \emph{KDD}.

\bibitem[{Borghol et~al.(2011)Borghol, Mitra, Ardon, Carlsson, Eager, and
  Mahanti}]{borghol_characterizing_2011}
Borghol, Y.; Mitra, S.; Ardon, S.; Carlsson, N.; Eager, D.; and Mahanti, A.
  2011.
\newblock Characterizing and modelling popularity of user-generated videos.
\newblock In \emph{Performance Evaluation}.

\bibitem[{Brodersen, Scellato, and Wattenhofer(2012)}]{brodersen_youtube_2012}
Brodersen, A.; Scellato, S.; and Wattenhofer, M. 2012.
\newblock {YouTube} around the world: geographic popularity of videos.
\newblock In \emph{WWW}.

\bibitem[{Bruns(2019)}]{bruns_after_2019}
Bruns, A. 2019.
\newblock After the ‘{APIcalypse}’: social media platforms and their fight
  against critical scholarly research.
\newblock In \emph{Information, Communication \& Society}.

\bibitem[{Bärtl(2018)}]{bartl_youtube_2018}
Bärtl, M. 2018.
\newblock {YouTube} channels, uploads and views: {A} statistical analysis of
  the past 10 years.
\newblock In \emph{Convergence}.

\bibitem[{Cha et~al.(2007)Cha, Kwak, Rodriguez, Ahn, and Moon}]{cha_i_2007}
Cha, M.; Kwak, H.; Rodriguez, P.; Ahn, Y.-Y.; and Moon, S. 2007.
\newblock I tube, you tube, everybody tubes: analyzing the world's largest user
  generated content video system.
\newblock In \emph{IMC}.

\bibitem[{Cheng, Dale, and Liu(2008)}]{cheng_statistics_2008}
Cheng, X.; Dale, C.; and Liu, J. 2008.
\newblock Statistics and {Social} {Network} of {YouTube} {Videos}.
\newblock In \emph{{International} {Workshop} on {Quality} of {Service}}.

\bibitem[{Chow(2018)}]{chow_how_2018}
Chow, A.~R. 2018.
\newblock How a {Bollywood} {Music} {Label} {Conquered} {YouTube}.
\newblock
  https://www.nytimes.com/2018/11/14/world/asia/t-series-youtube-india.html.
\newblock Accessed: 2021-04-06.

\bibitem[{Clifton and Mann(2011)}]{clifton_can_2011}
Clifton, A.; and Mann, C. 2011.
\newblock Can {YouTube} enhance student nurse learning?
\newblock In \emph{Nurse Education Today}.

\bibitem[{Figueiredo, Benevenuto, and Almeida(2011)}]{figueiredo_tube_2011}
Figueiredo, F.; Benevenuto, F.; and Almeida, J.~M. 2011.
\newblock The {Tube} over {Time}: {Characterizing} {Popularity} {Growth} of
  {Youtube} {Videos}.
\newblock In \emph{WSDM}.

\bibitem[{Finamore et~al.(2011)Finamore, Mellia, Munafò, Torres, and
  Rao}]{finamore_youtube_2011}
Finamore, A.; Mellia, M.; Munafò, M.~M.; Torres, R.; and Rao, S.~G. 2011.
\newblock {YouTube} everywhere: impact of device and infrastructure synergies
  on user experience.
\newblock In \emph{IMC}.

\bibitem[{Fisher and Taub(2019)}]{fisher_how_2019}
Fisher, M.; and Taub, A. 2019.
\newblock How {YouTube} {Radicalized} {Brazil}.
\newblock
  \url{https://www.nytimes.com/2019/08/11/world/americas/youtube-brazil.html}.
\newblock Accessed: 2021-04-06.

\bibitem[{Freeman and Chapman(2007)}]{freeman_is_2007}
Freeman, B.; and Chapman, S. 2007.
\newblock Is "{YouTube}" telling or selling you something? {Tobacco} content on
  the {YouTube} video-sharing website.
\newblock In \emph{Tobacco Control}.

\bibitem[{Gill et~al.(2007)Gill, Arlitt, Li, and Mahanti}]{gill_youtube_2007}
Gill, P.; Arlitt, M.; Li, Z.; and Mahanti, A. 2007.
\newblock Youtube traffic characterization: a view from the edge.
\newblock In \emph{IMC}.

\bibitem[{Gill et~al.(2008)Gill, Arlitt, Li, and
  Mahanti}]{gill_characterizing_2008}
Gill, P.; Arlitt, M.; Li, Z.; and Mahanti, A. 2008.
\newblock Characterizing user sessions on {YouTube}.
\newblock In \emph{Multimedia {Computing} and {Networking} 2008}.

\bibitem[{Horta~Ribeiro et~al.(2020)Horta~Ribeiro, Ottoni, West, Almeida, and
  Meira}]{hortaribeiroAuditingRadicalizationPathways2019}
Horta~Ribeiro, M.; Ottoni, R.; West, R.; Almeida, V. A.~F.; and Meira, W. 2020.
\newblock Auditing {Radicalization} {Pathways} on {YouTube}.
\newblock In \emph{{FAT}*}.

\bibitem[{Madathil et~al.(2015)Madathil, Rivera-Rodriguez, Greenstein, and
  Gramopadhye}]{madathil_healthcare_2015}
Madathil, K.~C.; Rivera-Rodriguez, A.~J.; Greenstein, J.~S.; and Gramopadhye,
  A.~K. 2015.
\newblock Healthcare information on {YouTube}: {A} systematic review.
\newblock In \emph{Health Informatics Journal}.

\bibitem[{Orsolic et~al.(2017)Orsolic, Pevec, Suznjevic, and
  Skorin-Kapov}]{orsolic_machine_2017}
Orsolic, I.; Pevec, D.; Suznjevic, M.; and Skorin-Kapov, L. 2017.
\newblock A machine learning approach to classifying {YouTube} {QoE} based on
  encrypted network traffic.
\newblock In \emph{Multimedia Tools and Applications}.

\bibitem[{Ottoni et~al.(2018)Ottoni, Cunha, Magno, Bernardina, Meira~Jr., and
  Almeida}]{ottoni_analyzing_2018}
Ottoni, R.; Cunha, E.; Magno, G.; Bernardina, P.; Meira~Jr., W.; and Almeida,
  V. 2018.
\newblock Analyzing {Right}-wing {YouTube} {Channels}: {Hate}, {Violence} and
  {Discrimination}.
\newblock In \emph{ACM WebScience}.

\bibitem[{Papadamou et~al.(2019)Papadamou, Papasavva, Zannettou, Blackburn,
  Kourtellis, Leontiadis, Stringhini, and
  Sirivianos}]{papadamou_disturbed_2019}
Papadamou, K.; Papasavva, A.; Zannettou, S.; Blackburn, J.; Kourtellis, N.;
  Leontiadis, I.; Stringhini, G.; and Sirivianos, M. 2019.
\newblock Disturbed {YouTube} for {Kids}: {Characterizing} and {Detecting}
  {Inappropriate} {Videos} {Targeting} {Young} {Children}.
\newblock In \emph{arXiv:1901.07046 [cs]}.

\bibitem[{{Pew Research}(2018)}]{pewresearchManyTurnYouTube2018}
{Pew Research}. 2018.
\newblock Many {Turn} to {YouTube} for {Children}'s {Content}, {News},
  {How}-{To} {Lessons}.
\newblock Pew Research Center: Internet, Science \& Tech.
\newblock Accessed: 2021-04-06.

\bibitem[{Rieder, Coromina, and
  Matamoros-Fernández(2020)}]{rieder_mapping_2020}
Rieder, B.; Coromina, O.; and Matamoros-Fernández, A. 2020.
\newblock Mapping {YouTube}.
\newblock In \emph{First Monday}.

\bibitem[{Schwind et~al.(2020)Schwind, Midoglu, Alay, Griwodz, and
  Wamser}]{schwind_dissecting_2020}
Schwind, A.; Midoglu, C.; Alay, O.; Griwodz, C.; and Wamser, F. 2020.
\newblock Dissecting the performance of {YouTube} video streaming in mobile
  networks.
\newblock In \emph{International Journal of Network Management}.

\bibitem[{Shuyo(2010)}]{shuyo2010language}
Shuyo, N. 2010.
\newblock Language detection library for java.
\newblock \url{https://code.google.com/archive/p/language-detection}.
\newblock Accessed: 2021-04-06.

\bibitem[{Siersdorfer et~al.(2010)Siersdorfer, Chelaru, Nejdl, and
  San~Pedro}]{siersdorfer_how_2010}
Siersdorfer, S.; Chelaru, S.; Nejdl, W.; and San~Pedro, J. 2010.
\newblock How useful are your comments? analyzing and predicting youtube
  comments and comment ratings.
\newblock In \emph{WWW}.

\bibitem[{Statista(2016)}]{statista_youtube_2016}
Statista. 2016.
\newblock Leading {YouTube} Markets 2016.
\newblock \url{https://bit.ly/34NAOV1}.
\newblock Accessed: 2021-04-06.

\bibitem[{Sureka et~al.(2010)Sureka, Kumaraguru, Goyal, and
  Chhabra}]{sureka_mining_2010}
Sureka, A.; Kumaraguru, P.; Goyal, A.; and Chhabra, S. 2010.
\newblock Mining {YouTube} to {Discover} {Extremist} {Videos}, {Users} and
  {Hidden} {Communities}.
\newblock In \emph{Information {Retrieval} {Technology}}.

\bibitem[{Wu, Rizoiu, and Xie(2018)}]{wu_beyond_2018}
Wu, S.; Rizoiu, M.-A.; and Xie, L. 2018.
\newblock Beyond {Views}: {Measuring} and {Predicting} {Engagement} in {Online}
  {Videos}.
\newblock In \emph{ICWSM}.

\bibitem[{Zhou et~al.(2011)Zhou, Li, Adhikari, and Zhang}]{zhou_counting_2011}
Zhou, J.; Li, Y.; Adhikari, V.~K.; and Zhang, Z.-L. 2011.
\newblock Counting {YouTube} videos via random prefix sampling.
\newblock In \emph{IMC}.

\bibitem[{Zhou, Khemmarat, and Gao(2010)}]{zhou_impact_2010}
Zhou, R.; Khemmarat, S.; and Gao, L. 2010.
\newblock The impact of {YouTube} recommendation system on video views.
\newblock In \emph{IMC}.

\bibitem[{Zink et~al.(2009)Zink, Suh, Gu, and
  Kurose}]{zink_characteristics_2009}
Zink, M.; Suh, K.; Gu, Y.; and Kurose, J. 2009.
\newblock Characteristics of {YouTube} network traffic at a campus network -
  {Measurements}, models, and implications.
\newblock In \emph{The International Journal of Computer and Telecommunications
  Networking}.

\end{thebibliography}
 }

\end{document}